\documentclass[a4paper,pre,reprint,twocolumn,10pt,superscriptaddress]{revtex4}
\pdfoutput=1
\usepackage[utf8]{inputenc}
\usepackage[english]{babel}
\usepackage{amsmath}
\usepackage{amsfonts}
\usepackage{amssymb}
\usepackage{graphicx}
\usepackage{color}
\usepackage{natbib}
\usepackage{textcomp}
\usepackage{hyperref}
\hypersetup{colorlinks=true,linkcolor=blue,citecolor=magenta,pdftitle={Plasma probe characteristics in  low density hydrogen pulsed plasmas}}

\newcommand*{\doi}[1]{\href{http://dx.doi.org/\detokenize{#1}}{doi:\detokenize {#1}}}

\begin{document}

 \title{Plasma probe characteristics in low density hydrogen pulsed plasmas}
 \author{D.I.~Astakhov}
 \affiliation{XUV Group, MESA+ Institute for Nanotechnology, University of Twente, P.O. Box 217, 7500 AE Enschede, The Netherlands}
 \author{W.J.~Goedheer}
 \affiliation{FOM Institute DIFFER - Dutch Institute for Fundamental Energy Research, P.O. Box 1207, 3430 BE Nieuwegein, The Netherlands}
 \author{C.J.~Lee}
 \affiliation{XUV Group, MESA+ Institute for Nanotechnology, University of Twente, P.O. Box 217, 7500 AE Enschede, The Netherlands}
 \author{V.V.~Ivanov}
 \author{V.M.~Krivtsun}
 \affiliation{Institute for Spectroscopy RAS (ISAN), Fizicheskaya 5, Troitsk 142190, Russian Federation}
 \author{A.I.~Zotovich}
 \affiliation{Faculty of Physics, Lomonosov Moscow State University, Leninskie Gory, Moscow 119991, Russian Federation}
 \affiliation{Skobeltsyn Institute of Nuclear Physics, Lomonosov Moscow State University, Leninskie Gory, Moscow 119991, Russian Federation}
 \author{S.M.~Zyryanov}
 \author{D.V.~Lopaev}
 \affiliation{Skobeltsyn Institute of Nuclear Physics, Lomonosov Moscow State University, Leninskie Gory, Moscow 119991, Russian Federation}
 \author{F.~Bijkerk}
 \affiliation{XUV Group, MESA+ Institute for Nanotechnology, University of Twente, P.O. Box 217, 7500 AE Enschede, The Netherlands}

\begin{abstract}
Probe theories are only applicable in the regime where the probe's perturbation of the plasma can be neglected. However, it is not always possible to know, \emph{a priori}, that a particular probe theory can be successfully applied, especially in low density plasmas. This is especially difficult in the case of transient, low density plasmas. Here, we applied probe diagnostics in combination with a 2D particle-in-cell model, to an experiment with a pulsed low density hydrogen plasma. The calculations took into account the full chamber geometry, including the plasma probe as an electrode in the chamber. It was found that the simulations reproduce the time evolution of the probe IV characteristics with good accuracy. The disagreement between the simulated and probe measured plasma density is attributed to the limited applicability of probe theory to measurements of low density pulsed plasmas. Indeed, in the case studied here, probe measurements would lead to a large overestimate of the plasma density. In contrast, the simulations of the plasma evolution and the probe characteristics do not suffer from such strict applicability limits. These studies show that probe theory cannot be justified through probe measurements.
\end{abstract}
\maketitle

\section{Introduction}
Low pressure ($1$..$100$~Pa) low density \hbox{($n_e\sim10^8..10^9$~/cm$^3$)} pulsed plasmas are commonly found in many laboratory experiments. These plasmas can exhibit complicated behavior because they can be operated in the non-local and non-stationary regime. 

However, in practice, plasmas of this type are frequently investigated with the Langmuir probe technique. But the application of plasma probe diagnostics is hindered, because it strictly requires \cite{Hutchinson.2005.principles,Chen.2003.langmuir}  (a) that the plasma is distorted by the probe only locally in a well defined manner, and (b) that the evolution of the plasma, excluding the distorted region, is not affected by the presence of the probe. 

Probe diagnostics were initially designed for glow discharge plasmas, where the applicability limits could be estimated in advance easily. Thus, Langmuir probe measurements allowed plasma parameters, such as the electron density and temperature, to be derived with the help of appropriate formulas. However, the probe method does not provide a technique to estimate the applicability of the method itself.

Nevertheless, in many experiments with a lower density plasma there is not enough information to justify the probe method beforehand. Thus, the plasma parameters, which are obtained from probe measurements with help of probe theory, are used to evaluate the methods applicability. 

Applying a probe theory outside of its applicability limits leads to errors in the plasma parameters derived from the measured probe response. Frequently, in day to day practice, the probe method validity is estimated from two observations: (a) the probe I~--~V characteristic should have distinct ion and electron contributions, and (b) the plasma size should be at least a factor of 100 larger than the estimated Debye length \cite{Chen.2003.langmuir}.  However, these parameters, derived from the probe measurements, have a tendency to support the validity of the probe method, e.g. if the derived electron density is overestimated for a low density plasma, the second condition is satisfied in most cases. 

In order to overcome these difficulties, we propose to use a combined approach, where experiments and simulations are coupled as tightly as possible, with the full experimental geometry included in simulations. In these simulations, the plasma probe is included as an additional electrode. By including the probe in the model, any effect that it has on the plasma is naturally included in the dynamics. This approach does not have any of the restrictions discussed above. 

In general this requires a 3D plasma model. Such simulations are extremely time consuming, but, fortunately, it is possible to reduce the model dimensions by using of an axis-symmetrical configuration for the experimental setup. This allows  a 2D cylindrically symmetric model to be used for the plasma simulation, which is much faster than a 3D model. 

In the present work, we consider a low pressure pulsed hydrogen plasma. Because hydrogen is the lightest element, we can expect the fastest response from ions.

Good agreement was obtained between measured and simulated probe characteristics for several conditions. Thus, the  plasma parameters obtained from the simulations, such as electron and ion densities, and temperatures, are expected to be close to the experimental values. However, the simulated plasma parameters differ from the plasma parameters derived from probe theory calculations, based on the measured probe response. This difference was attributed to the distortion of the plasma by the probe, and to transient plasma dynamics.

\section{Experimental setup \label{section:experiments}}
\begin{figure}
    \includegraphics[width=\columnwidth]{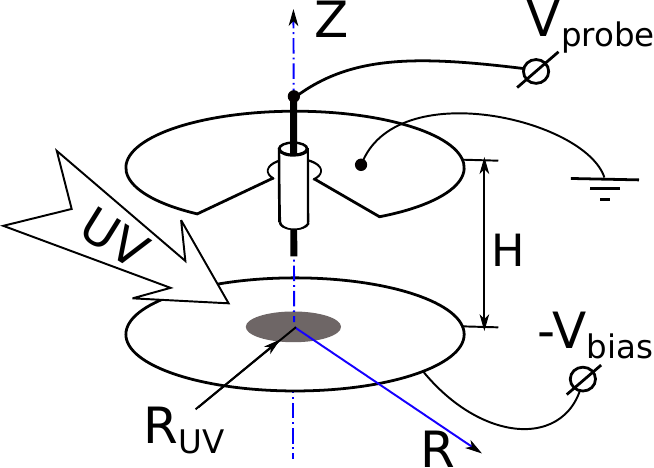}
    \caption{Configuration of the experimental chamber. The electrodes are 10~cm in diameter and separated by 5~cm. The plasma probe passes through the center of the top electrode and is 3~cm from the bottom electrode. The open end of the probe is 1~cm in length and 0.5~mm in diameter. The remainder of the probe is covered by a dielectric with a length of 1~cm and a diameter of 1.3~mm. The plate holders and chamber walls are not show, because  the vacuum chamber is large (cylinder 20~cm in diameter and 20~cm in height). \label{fig:uv_chamber}}
\end{figure}

The interior configuration of the vacuum chamber, in which the low density H$_2$ pulsed plasma was generated, is presented in Fig.~\ref{fig:uv_chamber}. The chamber was pumped with a turbomolecular pump to keep the residual gas pressure lower than 0.003~Pa. 
During experiments, hydrogen flowed through the chamber at several sccm, controlled by a gas flow meter. The hydrogen pressure in the chamber was varied in the range of 0--60~Pa.
  
The aluminum electrodes are  circular plates  with a 10~cm diameter. The distance between the electrodes can be varied in the range of 2--5~cm. In the experiments, the bottom electrode was negatively biased at -200--0~V, while the  top electrode was grounded. The plasma was ignited by the photoelectrons emitted from the bottom electrode during the UV laser pulse. Most of these electrons are reflected back to the bottom electrode by the space charge field. The electrons which pass the potential barrier, are accelerated by the applied bias field and ionize the background gas, thus igniting the plasma.

A KrF excimer laser (LPX210) was used as the source of the UV radiation that produces photoelectrons at the bottom electrode. The temporal profile of the laser pulse consists of a short pulse with a full width half maximum (FWHM) of $<$50~ns and a tail, extending up to 200--300~ns in duration (see Fig.~\ref{fig:vacuum_current}).  The pulse energy, incident  on the bottom electrode, was approximately 10~mJ/pulse, which was only 20$\%$ of total laser pulse energy. Approximately the same fraction  of the total laser energy was reflected to a detector by a beam splitter to control the laser pulse energy during the experiment. The laser beam was passed through a diaphragm to obtain a uniform intensity distribution with a diameter of 1.3~cm at the bottom electrode. The laser light was incident at an angle to the electrode after entering the chamber through a quartz window.

The plasma probe was installed along the  symmetry axis of the chamber. The probe  was  0.05~cm in diameter and 1~cm long, the glass shield that covers the rest of the probe was 0.13~cm in diameter. The probe measurements were automated with a NI-DAQ card and specially developed signal processing unit (SPU), consisting of a trans-impedance isolated amplifier, differential amplifier, and probe bias control. This unit allows probe currents down to 10~nA to be measured in the band of approximately 6~MHz. The signals amplified by the SPU were transferred to an oscilloscope (Tektronix TDS3032). The overall control of measurements and the experimental setup was performed with a computer program via the NI-DAQ card, as well as GPIB, and RS232 interfaces. The obtained data were stored on a server for further analysis.

\section{Model \label{section:model}}

The plasma was modelled using a two-dimensional ($rz$ cylindrically symmetric) particle-in-cell (PIC) model. The plasma species include electrons, H$^+$,H$_2^+$ and H$_3^+$ as particles for accurate description of processes in the plasma sheath. Our model follows the general PIC scheme, described elsewhere \cite{Birdsall.1985.plasma}. 

The combination of measurements and simulations of discharge characteristics  and plasma probe measurements allows the number of free parameters to be reduced. These parameters (e.g. ion induced electron emission yield, photo electron yield, initial electron energy distribution etc.) are  required to perform  simulations, but the range of values  found in literature are typically  too large for accurate modeling. It is worth noting here that many such parameters are intrinsically setup dependent (i.e. the photo electron yield could be very sensitive to the surface conditions). Thus, preferably, such parameters should be measured.

An accurate measurement of all unknown parameters is often unfeasible. To solve this issue, the model was developed in parallel to the experimental setup. Thus, many of the unknown parameters were obtained with reasonable accuracy  using the results from simple experiments, which were performed as a test procedure during the construction and calibration of the experiment.

\subsubsection{UV laser and electron spectrum used as input to the model}
In the experiment, an excimer laser, operating at a wavelength of 248~nm, generates photoelectrons from the bottom electrode. In the model, this process is included as a photoelectron source on the bottom electrode. On each time step, when the laser is active, photoelectrons are released from the irradiated part of the bottom electrode to the computational domain. The amount of injected electrons is linearly proportional to the instantaneous laser intensity. The electron energies are sampled according to a  distribution, as discussed below. The velocities of the injected electrons  are distributed according to a cosine law relative to the normal direction to the boundary.

The initial electron energy distribution and the temporal profile of the  laser pulse are  important parameters  that significantly influence the plasma that is formed later. Under vacuum conditions, where no plasma is ignited, the temporal profile of the current contains information about the laser pulse shape and the effective photoelectron yield. We used this experiment and the following physically motivated choice of parameters to define the effective electron source for the model.

The laser energy per pulse in the experiment was 10~mJ. The  typical photoelectron emission yield is estimated to be 10$^{-4}$--10$^{-3}$~electrons/photon \cite{Raizer.1991.gas},
thus, the maximum collected charge is approximately 20--200~nC. However, in the experiment, the collected charge per pulse under vacuum conditions was approximately 0.2~nC. Hence,  the space charge current limiting was very high during the main part of the  laser pulse.

The space charge potential barrier formation leads to a locally high electron density. The electrons, which are trapped by the barrier, have small energies, thus, Coulomb collision processes should dominate. From a computational point of view it is  not reasonable to compute the full collision kinetics of the cold electrons, since it is known to lead to the formation of a Maxwellian tail in the energy distribution, and the detailed kinetics depend on the exact initial distribution and photoelectron yield, which are not known. Thus, a Maxwellian function was taken as the initial electron energy distribution function (EEDF).

The effective temperature for this EEDF was heuristically chosen from two observations. Firstly, the bottom electrode was made from aluminum, which might by partially oxidized. A number of values for the work function of  this material have been reported, the lowest of which is about~4~eV \cite{Lide.2003.crc}. Secondly, since the photoelectron yield is not known precisely, we only need to find a reasonable combination of effective temperature and yield to reproduce the charge-bias characteristic, as measured experimentally under vacuum conditions. Thus, the effective temperature for EEDF was chosen to be 0.5~eV, and the value for photoelectron  yield was fine-tuned from the measured vacuum charge bias characteristics of the system. Because of the significant space charge screening, the sensitivity for the particular choice is small. 

A scan over laser pulse energies from 3~mJ to 15~mJ was performed to ensurethat the experiment was stable. It was found that the vacuum discharge characteristics have the lowest sensitivity to the laser power variation near 10~mJ/pulse. However, while the integral value was stable, there were signs of higher order processes (e.g. two photon photoelectric effect), since the value of the vacuum current peak changed near a laser pulse energy of 6~mJ (see Fig.~\ref{fig:vacuum_current}). These experiments were included into the fitting procedure, and it was found that while for all energies below 6~mJ/pulse the above described procedure produced consistent results, it  was not possible to reproduce the current peak value for 10~mJ/pulse and above due to space charge screening. For this reason, a very small amount of 4~eV electrons, with a yield of approximately $10^{-8}$, was added to the electron source to describe the case of 10~mJ/pulse laser energy.

As mentioned above,  the laser pulse shape has a tail (up to 300~ns) in comparison to the 60~ns main pulse (See  Fig.~\ref{fig:vacuum_current}).  This tail plays a very important role in plasma formation as discussed in section \ref{section:discharge_characteristics}. Unfortunately, the  radiation intensity in the laser tail was too low to accurately measure with our equipment. This fact also suggests that one should not expect a space charge effect. For this reason, the EEDF for the electrons generated by the tail of the pulse is constant for all electron energies less than 1~eV, and zero for higher energies.

A fitting procedure was used to derive the laser tail shape from the measured discharge characteristics in vacuum. During this procedure the pulse shape was represented as the sum of the measured laser pulse  and a tail. The electron yield was chosen such that it reproduced the  peak current value and the  tail current simultaneously.

\subsubsection{Ion induced electron emission}

It is important to take into account ion induced electron emission, since free electrons are accelerated by the electric field to the top electrode and produce additional electrons and ions through impact ionization. The electrons produced through impact ionization contribute significantly to the later dynamics of the plasma and should be taken into account.

The bottom electrode in the experiment was biased at -200 --- 0V, therefore, the maximum ion energy is about 200~eV. For this energy range, the ion induced electron emission can be represented as a sum of potential emission (PE) and kinetic-induced emission (KE) \cite{Lakits.1990.threshold,Winter.1991.recent}. The PE yield is determined by the combination of the ion and target material, however, for the slow ions it is independent of the ion velocity. But, the KE yield (see below) strongly depends on the ion velocity.

In the model, the PE probability to emit an electron per one ion impact at the bottom electrode is computed with the following semi-empirical formula from~\cite{Kishinevsky.1973.estimation} 
\begin{equation}
    \gamma \approx \frac{0.2}{E_F} (0.8 E_i - 2 W)
    \label{eqn:potential_emission_coeff}
\end{equation}
Here $E_F$ is the metal Fermi energy, $E_i$ is \textit{ionization} energy of the ion and $W$ is the metal's work function. The applicability range of Eqn.~\ref{eqn:potential_emission_coeff} is $3W \leq E_i \leq (2E_F + W)$, the accuracy of this approximation in these range is not better than 10$\%$. Outside this range, the accuracy of Eqn.~\ref{eqn:potential_emission_coeff} slowly decreases.

The top and bottom electrodes (see Fig.~\ref{fig:uv_chamber}) are made from aluminum. The tabulated values for aluminum are $E_F = 11.7$~eV, $W$ = 3.8--4.5~eV. The large spread in work function can be attributed to a combination of the effect of the  surface crystalline structure \cite{Lide.2003.crc} and of surface oxide, which can lead either to an increase or a decrease of the observed work function, depending on the exact conditions \cite{Agarwala.1974.work}. The plasma probe is made from molybdenum, which has a Fermi energy and work function of $E_F$ = 6.8~eV, $W$ = 4.2--4.6~eV~\cite{Lide.2003.crc,Sosa.2002.electron}. 

For a hydrogen plasma there are three types of ions possible. The ions have ionization energies of: H$^+$ 13.6~eV, H$_2^+$ 15.4~eV, and H$_3^+$ 3.6~eV. The value used for H$_3^+$ here is the ionization potential for a long lived H$_3$ taken from \cite{Helm.1988.measurement}. This definition of H$_3^+$ ionization energy is less than twice the energy of the work functions of aluminum or molybdenum, thus, H$_3^+$ does not produce any potential emission. Hence, the PE probability for H$^+$ is approximately $0.04 \pm 0.02$ on aluminum and $0.06 \pm 0.02$   on molybdenum, while for H$_2^+$ ---  $0.07 \pm 0.02$ on aluminum and  $0.1  \pm 0.02$ on molybdenum. In these estimates we included an error of at least 10$\%$ due to the approximation used and  further 10$\%$ error due to the poor knowledge of work function.

There are complicated KE models \cite{Winter.2001.slow-ion}, but they require some fitting parameters to be defined. Since the exact surface conditions are not known well, the same accuracy is obtained by simply fitting to experimental data. Therefore, we follow the same approach as in \cite{Bogaerts.2002.hybrid}, and use a fit to the experimental data~\cite{Winter.1991.recent} to determine for KE for all metallic surfaces.
\begin{equation}
 \gamma_{KE} \approx 6.2\cdot10^{-5} \cdot (E\mbox{[eV/amu]})^{1.15}
\end{equation}
Here $E$ is the \textit{kinetic} energy of the ion per ion mass.

The physical motivation for this approach is as follows. For many metals and conductive materials, the work function ($W$) is approximately 5~eV and the quasi-classical threshold  of KE is $v_i > W/(2k_F)$, here $v_i$ is the ion velocity and $k_F$ is a Fermi impulse. KE depends strongly on the ion type and target material, especially for heavy ions. But, in experiments~\cite{Winter.1991.recent}, the KE  probability, as a function of velocity, varies only by a factor of two or less for H$^+$, H$_2^+$ and H$_3^+$ (for a gold target). Similar experiments indicate that, for a given ion species, KE is similar for a broader range of target materials~\cite{Baragiola.1979.electron}.

In our simulation, the use of an ion-type-dependent ion-induced electron emission coefficient leads to  time-dependent ion-induced emission, because, just after the UV pulse, the main ion is H$_2^+$ and the ion induced emission coefficient is large. However, due to a very efficient ion conversion reaction  \mbox{H$_2^+$ + H$_2$ $\rightarrow$ H$_3^+$ + H}, the role of ion-induced emission significantly decreases after a characteristic ion conversion time.

Although the approximations for PE and KE are rather simple, they provide reasonable estimates of ion-induced emission. For instance, this simple model allows us to distinguish between a discharge initiated by a laser pulse without a tail and a large ion-induced emission coefficient, and the case with a smaller ion-induced emission coefficient and a laser pulse with a tail (as in Fig.~\ref{fig:discharge_current}).

\subsubsection{length scales, grid resolution}
The spatial resolution of the rectilinear grid is chosen to resolve two main characteristic length scales: the space charge potential well and the Debye length. The space charge potential well length scale can be estimated from the analytical solution for a 1D diode with a Maxwellian initial electron energy distribution \cite{Langmuir.1923.effect}. In the considered case, most photoelectrons are reflected back to the surface. Thus, it is possible to simplify the formulas from \cite{Langmuir.1923.effect} and obtain
\begin{equation}
   z_m \simeq 0.1\text{ cm} \times \frac{(T[\text{eV}])^{3/4}}{(I[\text{mA/cm$^2$}])^{1/2}}
\end{equation}
Here, $T$ is the initial temperature of emitted photoelectrons, $I$ is the current density near the cathode and $z_m$ is the distance from the cathode and the bottom of the space charge potential well. For $T\sim 0.5$eV and $I \sim 2 $mA/cm$^2$, one obtains  $z_m \sim$ 0.04~cm. Since the UV spot diameter is much larger than the estimated $z_m$, the $z_m$ length scale should only be resolved in the $z$ direction near the bottom electrode. For the plasma region, the grid cell size is estimated as the  Debye radius, $r_{D}$, for a plasma with $n_e \sim 10^9$~/cm$^3$, and $T_e\sim0.5$~eV, which corresponds to $r_D \sim$ 0.02~cm. We ran several tests to ensure that the grid resolution does not affect our results.

The displacement currents to the probe and electrodes are also included in the model. They are computed at each time step as the numerical derivative of the electric field over each of the electrode boundaries. Inclusion of the displacement current allows direct comparison between the simulated and experimental discharge currents.

To calculate the production of the different ion species, a set of the cross-sections for the hydrogen chemistry was assembled from the several sources. The data 
from \cite{Phelps.1990.cross} was used as a base, and extended using the work of \cite{Simko.1997.transport}, \cite{Mokrov.2008.monte_carlo}. 

The procedure described in \cite{Nanbu.1994.simple} is used to perform Monte Carlo
collisions with the background gas. This procedure allows  a reaction channel to be chosen before sampling the reaction probability, which offers significant  performance improvement compared to the Null collision scheme~\cite{Skullerud.1968.stochastic}.

We tested the consistency of our implementation by modeling  swarm experiments and found good agreement with experimental values \cite{Dutton.1975.survey} for electron ionization coefficients, electron mobility and for H$^+$, H$_3^+$ mobility in hydrogen \cite{Graham.1973.mobilities}.

\section{Results and discussion\label{section:results}}

\subsection{discharge characteristics \label{section:discharge_characteristics}}
Let us begin with the analysis of the discharge characteristics of the experimental setup without a plasma probe. 

During the laser pulse, only a small amount of photoelectrons enter the chamber because of the space-charge-induced potential barrier. This barrier is not just due to the electron cloud trapped near the bottom electrode, but is built up by all the excess negative charge present in the chamber. Hence, simulations under vacuum conditions (i.e., no plasma) provide a good estimate of the temporal profile of the laser pulse, but they are not very sensitive to the magnitude of electron emission. The laser pulse shape was fine-tuned such that it reproduced the current maximum in the case of 30~Pa, -200~V bias. This was sufficient to yield good agreement with experimental data for the maximum current under all other considered conditions. 

\begin{figure}      
      \includegraphics[width=\columnwidth]{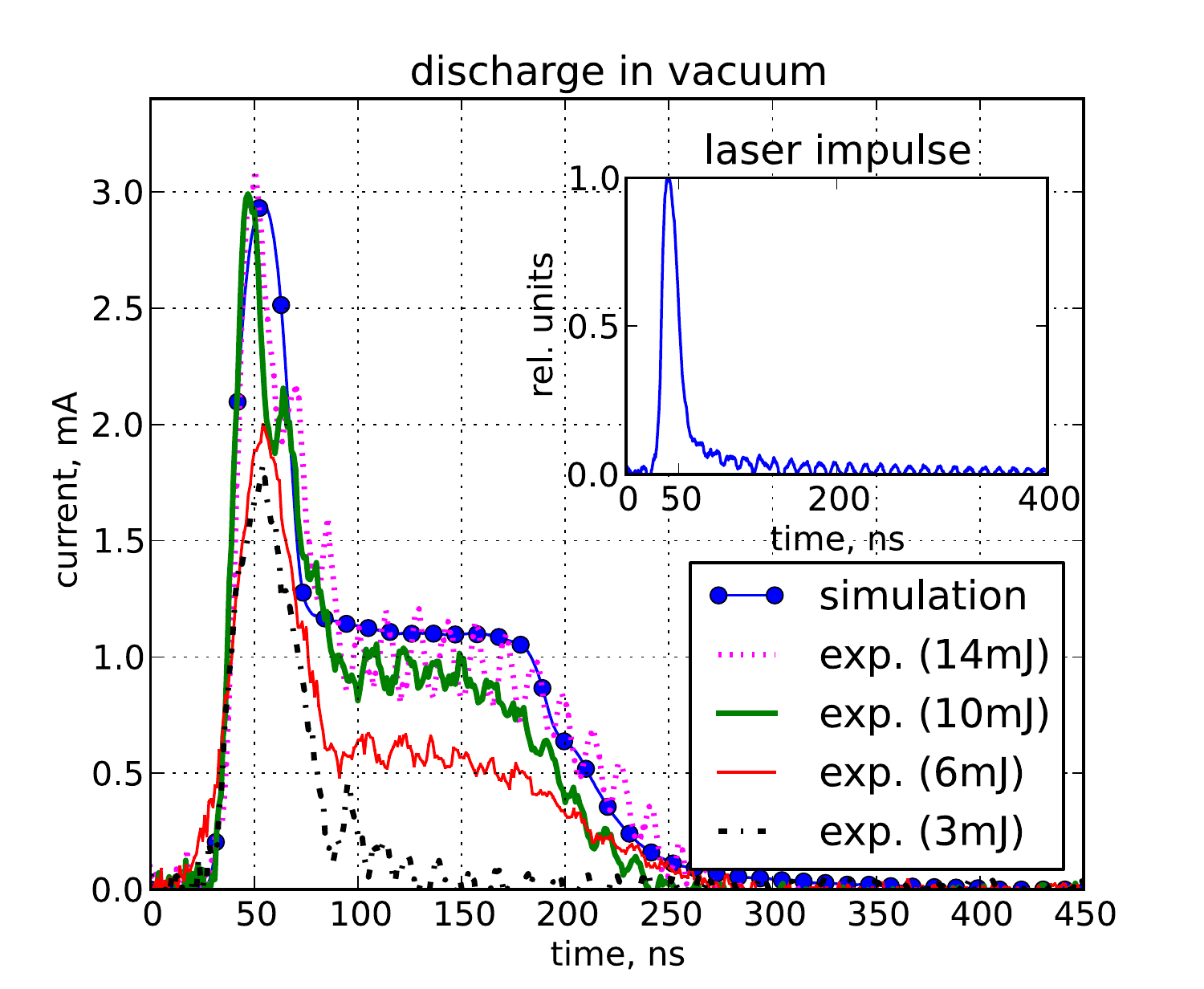}
      \caption{Discharge current in vacuum conditions, -200V bias. Laser pulse temporal profile for 10~mJ/pulse (inset). The current due to the main laser pulse is in the saturation regime. The difference between 6~mJ/pulse and 10~mJ/pulse peak currents can be attributed to some higher order effects, since the peak values for 3~mJ/pulse and 6~mJ/pulse are very close. Step-like current tail is due to laser tail. The current tails for 10mJ/pulse and 14~mJ/pulse are very close in magnitude,  justifying the choice of 10~mJ/pulse as the energy for experiments. The temporal  pulse shape of the laser, as measured by the detector is presented in the insert. \label{fig:vacuum_current}} 
\end{figure}
\begin{figure}
      \includegraphics[width=\columnwidth]{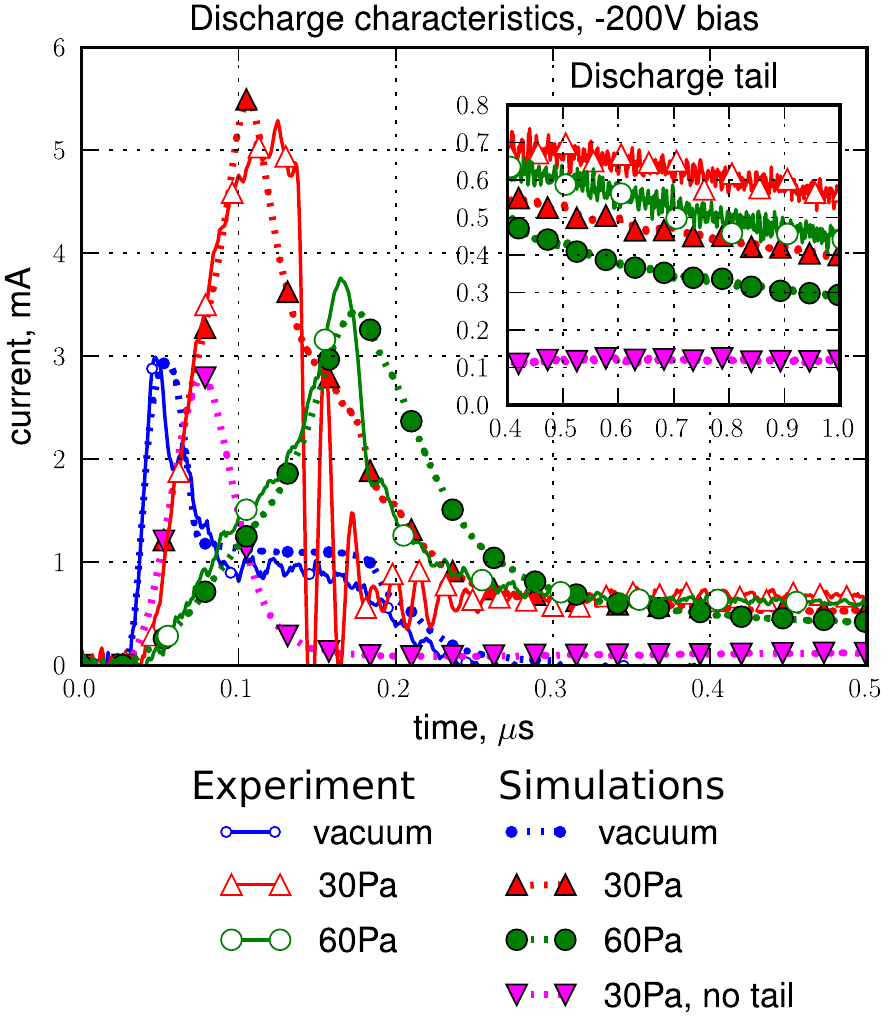}
      \caption{
      Discharge current as function of time for various hydrogen pressures. The tail of discharge is shown in inset in the same units as main plot. Note, that the laser pulse tail (see Fig.~\ref{fig:vacuum_current}) produces a significant contribution to plasma formation, as suggested by comparison of simulation results for 30~Pa with and without laser pulse tail.
      \label{fig:discharge_current}} 
\end{figure}
The current as function of time for -200~V bias at different background pressures is presented in Fig.~\ref{fig:discharge_current}. The laser pulse tail (see Fig.~\ref{fig:vacuum_current}) produces a significant contribution to the plasma formation (see comparison of cases with and without tail in Fig.~\ref{fig:discharge_current}). This happens because the electrons produced during the body of the laser pulse provide ionization in the chamber, which leads to the formation of the plasma. The presence of this plasma changes the potential distribution inside the chamber, leading to an increased field strength near the bottom electrode, which, in turn, increases the efficiency of ionization, thus, enhancing the effect of the laser pulse tail.

Despite reproducing both the slope of the current increase, and the current maximums for both pressures, the magnitude of the simulated discharge tail is about $30\%$ lower than in the experiment. Agreement between our simulations of the discharge tail, and the experimental data can be obtained by varying the tail of the laser pulse. Under these conditions, however, the peak position and width of the current were significantly distorted at pressures of 30~Pa and 60~Pa. Alternatively, increasing PE and KE also yields a discharge tail in agreement with experimental data. This, however, results in unrealistically high ion induced electron emission yield for H$_3^+$ ion (e.g. $\sim0.05$~electron/ion on average). Moreover, the model fails to reproduce the Langmuir probe measurements in this case.

Therefore, the disagreement between the model and experimental data is probably due to an accumulation of minor differences between the input parameters of the model and the experiment. These could be, for example, the laser spot size, uncertainties in the cross-sections set, pulse-to-pulse variations of laser pulse tail, etc.. Nevertheless, the $30\%$ difference between the plasma model and the experimental data is reasonably accurate for the probe simulations.

\subsection{plasma scalability}
\begin{figure}[h]
    \includegraphics[width=\columnwidth]{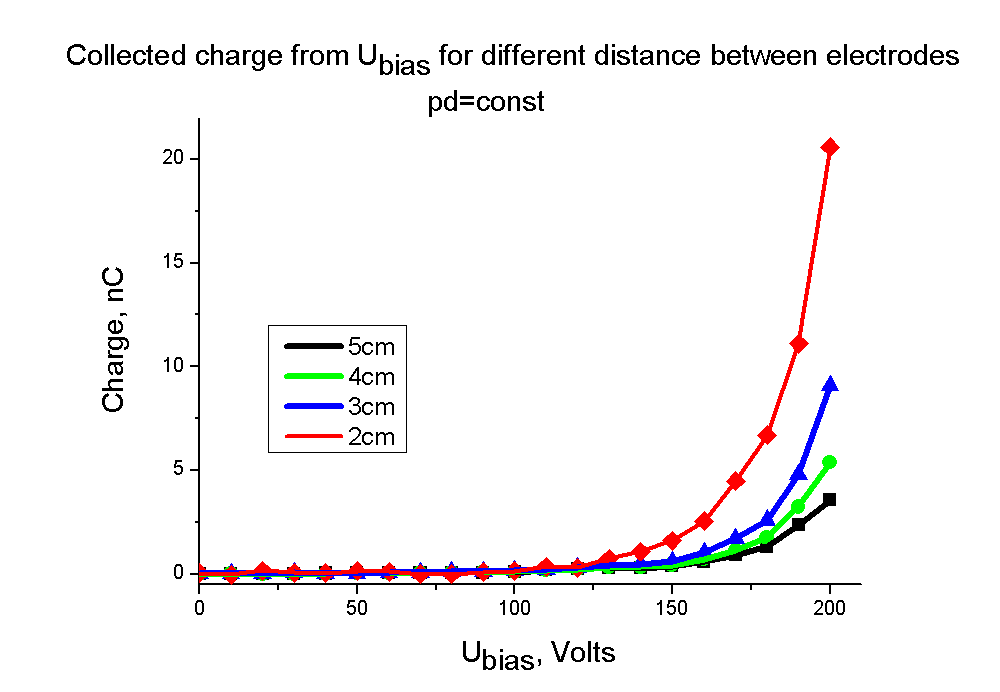}
    \caption{Collected charge for different distances between electrodes, but constant $pd$ (for $d=5$cm, $p = 30$Pa). \label{fig:pd_scaling_charge} }
\end{figure}

Additional experiments were performed to study the plasma scalability for different distances between electrodes, with fixed $pd$ parameter (here is $p$ --- hydrogen pressure, $d$ is the distance between electrodes) and fixed reduced electric field strength $E/p = U_{bias}/(pd)$.  The integral of the discharge current for different distances between the electrodes is presented in Fig.~\ref{fig:pd_scaling_charge}. The obtained plasma did not scale with $pd$ parameter. It showed that dynamic effects were important in plasma generation.

\subsection{Axially symmetric plasma probe}
\begin{figure}[h!]
    
    \includegraphics[width=0.5\textwidth]{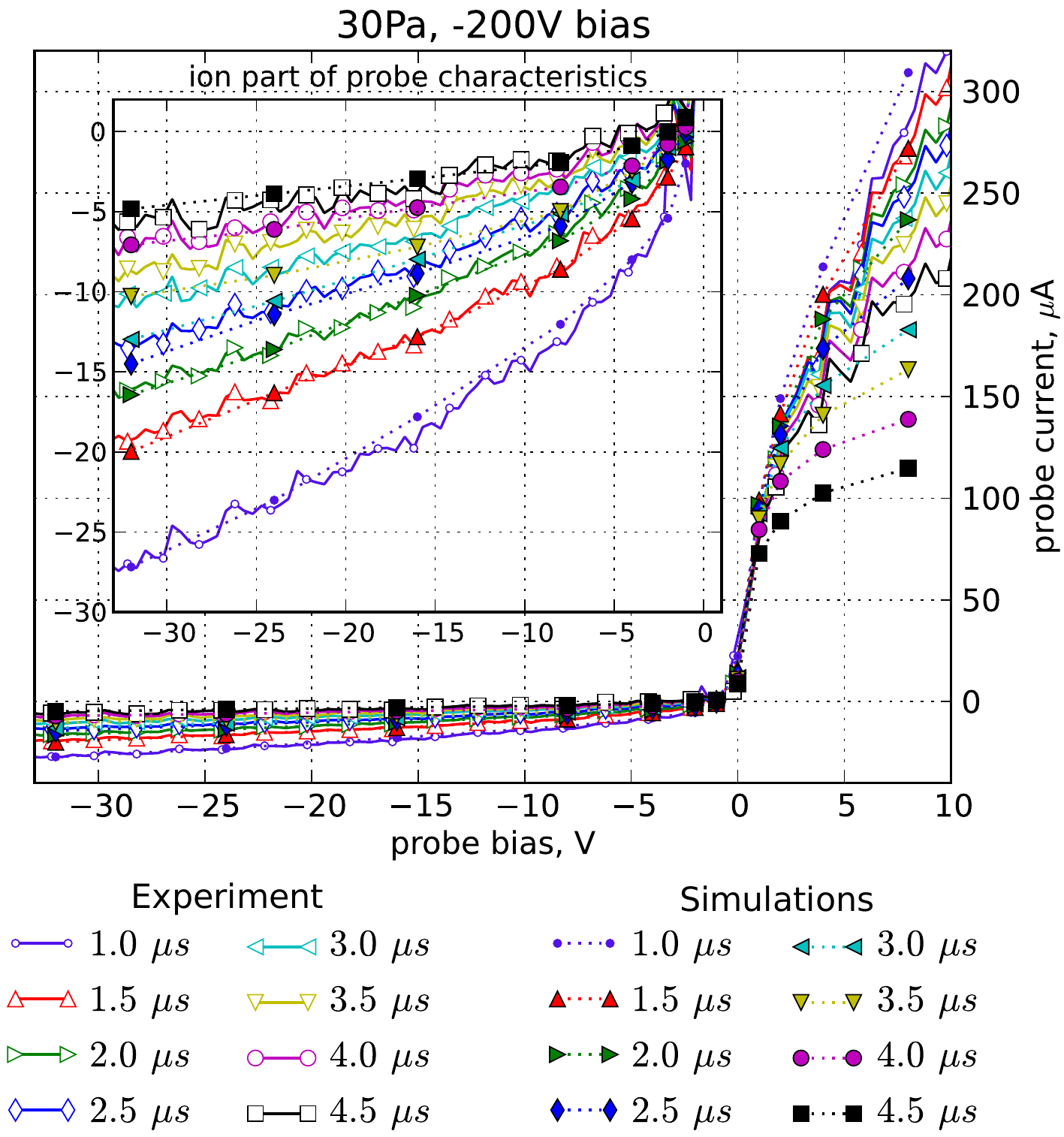}
    \caption{Comparison between  simulated and measured probe characteristic for 30~Pa and -200~V bias. The ion part of the probe (scales: $\mu$A vs V) characteristic is shown in the inset. \label{fig:probe_characteristic_30Pa}}  
    
    \includegraphics[width=0.5\textwidth]{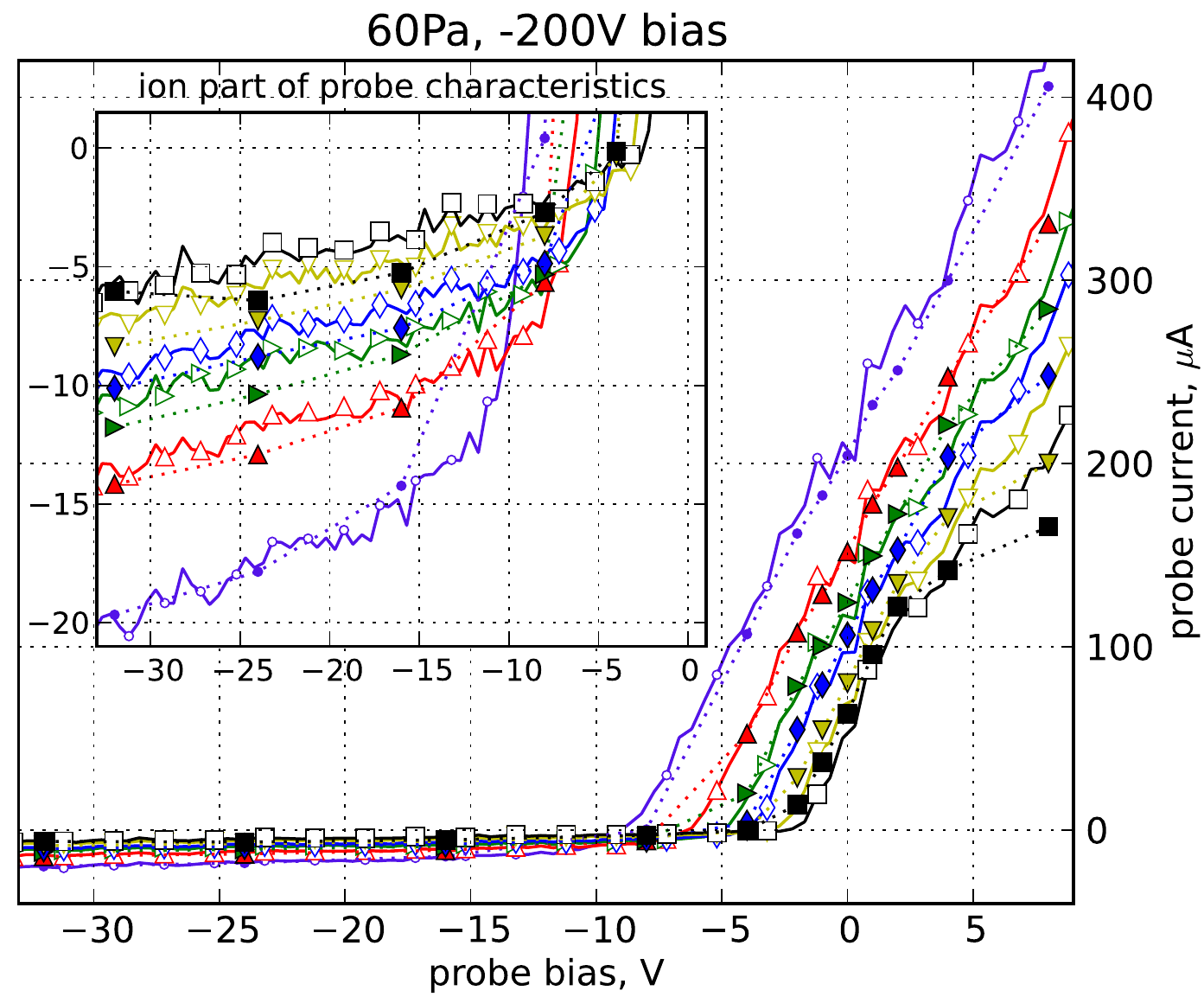}
    \caption{ Comparison between simulated and measured probe characteristic  for 60~Pa and -200~V bias. The ion part of probe (scales: $\mu$A vs V) characteristic is shown in the inset. The line legend is the same as in Fig.~\ref{fig:probe_characteristic_30Pa}. 
    \label{fig:probe_characteristic_60Pa} }
\end{figure}

To make a detailed investigation of plasma evolution the plasma probe diagnostic was used. The probe was installed on the symmetry axis of the chamber to allow direct simulations of the probe characteristics with the 2D plasma model. 

The probe measurements were performed at 30~Pa and 60~Pa hydrogen, the bias between the electrodes was -200~V in both cases. The simulated and measured probe characteristics were shifted by 1~V for 30~Pa and by 2~V for 60~Pa. This shift could have numerous causes, therefore, the measured probe characteristics were shifted by these values. The measured probe characteristics and the comparison with the simulations are presented in Fig.~\ref{fig:probe_characteristic_30Pa} and Fig.~\ref{fig:probe_characteristic_60Pa}. All presented simulations use the same initial input data apart from those varied in the experiments. 

To take into account the characteristic response time of the SPU, the simulated probe response was numerically convoluted with $C \cdot exp(-t/t_0)$, here $t_0\sim100$~ns is the SPU response time and constant $C$ is chosen such that the collected charge is preserved during numerical integration.

The agreement between the simulated and measured probe characteristic is good up to 2.5$\mu$s after the start of the UV pulse. At later times, the accuracy decreases, due to the response of the simulated probe changing because the tip of the probe is no longer in the region containing a plasma (see Fig.~\ref{fig:plasma_with_probe}). The transition of the probe from inside to outside the plasma occurs between 2.5~$\mu$s and 3.5~$\mu$s, however, the exact transition time is very sensitive to the model input parameters. After the transition (4.0-4.5~$\mu$s, 30~Pa, -200~V), there is closer agreement between the measured and modelled ion contribution to the probe characteristic. Furthermore, despite the plasma no longer covering the entire probe, the probe response appears normal. 

For the case of 60~Pa, -200~V (Fig.~\ref{fig:probe_characteristic_60Pa}), the transition occurs later due to a larger contribution from collisions with the background gas.
Therefore, the ion contribution to the probe characteristic for 3.5~$\mu$s in Fig.~\ref{fig:probe_characteristic_60Pa}, while still distorted, are close to the experimental ones.

The electron contribution to the simulated probe characteristics in Fig.~\ref{fig:probe_characteristic_30Pa} deviates significantly from the experiment occasionally from 2~$\mu$s onwards for a pressure of 30~Pa. But, at 60~Pa (Fig.~\ref{fig:probe_characteristic_60Pa}), the deviation of the electron part is much smaller and only pronounced from 3.5~$\mu$s and for a probe bias larger than 5~V. These differences can be attributed to the combined effects of the uncertainties in electron~--~neutral cross-section set and some secondary effects (e.g. reflection of electrons from the aluminium surface), which are not taken into account. However, the simulations accurately reproduce the transition from the ion part to the electron part of the probe characteristic in both cases. 

\begin{figure}[!h]
     \includegraphics[width=\columnwidth]{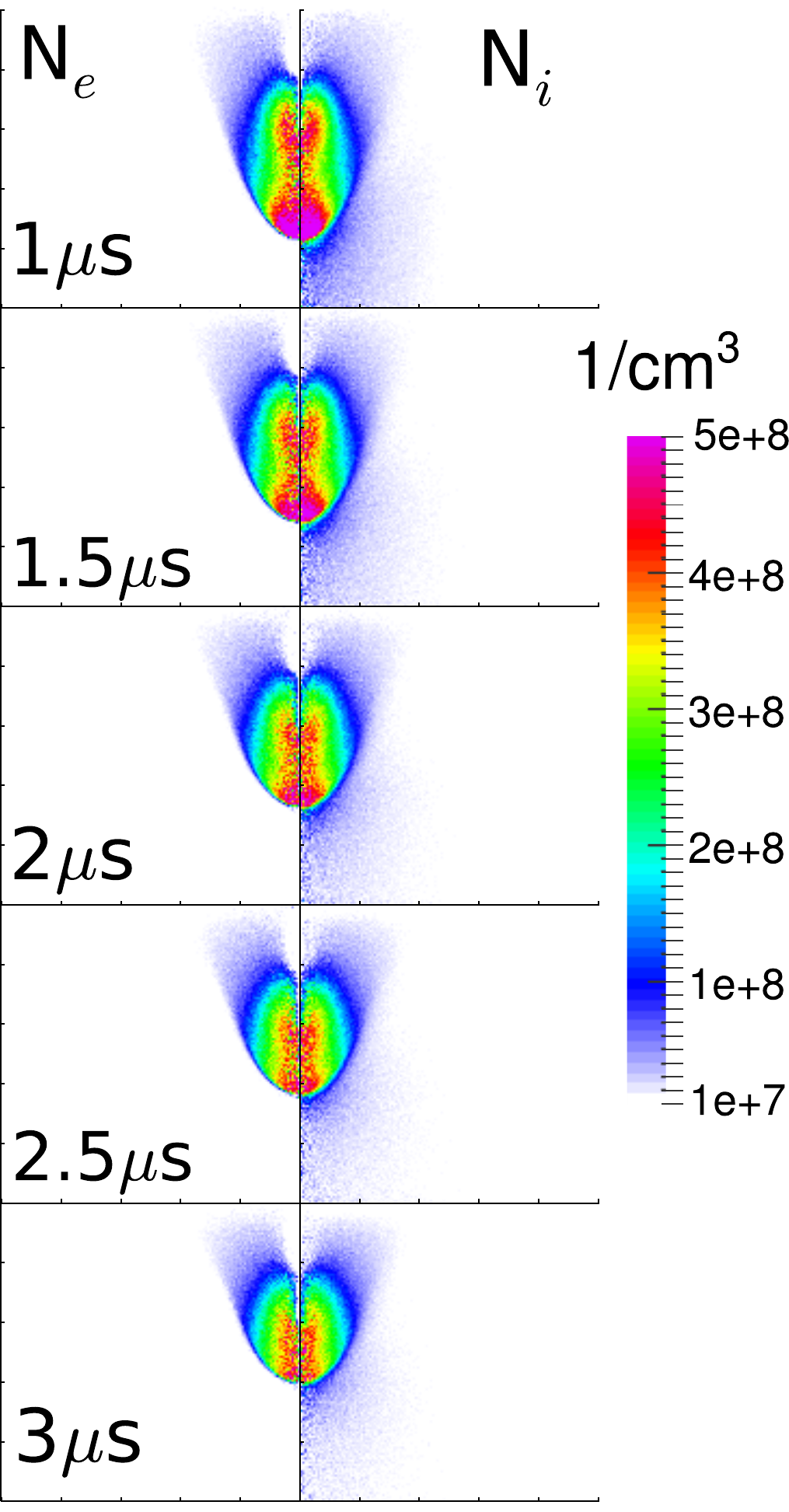}
     \caption{Electrons and ions (summed over ions types) densities for the different moments of time from UV pulse start. Simulation for 60~Pa H$_2$, -200~V bias between electrodes and -4~V on the probe. The probe current for this conditions change sign from positive at 1.5~$\mu$s to negative at 3~$\mu$s, inspite of the probe being completely covered by the plasma.  \label{fig:plasma_with_probe_60Pa}}
\end{figure}

Interestingly, the apparent shift of the probe characteristic in Fig.~\ref{fig:probe_characteristic_60Pa} with time is reproduced  by the model. Usually such a shift in the stationary plasma conditions may be explained by partial presence of the probe in or near the plasma sheath. However the reason of the shift in the case of  a pulsed plasma may have other causes. As is seen in Fig.~\ref{fig:plasma_with_probe_60Pa}, the change of the sign of the probe current can not be explained by the presence of the probe in the sheath.

\subsubsection{Comparison between computed plasma density and that derived from measured probe characteristic}
We applied the  Bernstein-Rabinowitz-Laframboise (BRL) \cite{Laframboise.1966.theory,Chen.2003.langmuir}, theory to analyze the probe I~--~V characteristics. We used the ESP\_BRL program \cite{Chen.2003.langmuir} to obtain BRL fit for the probe characteristics. The fit results are presented in Table.\ref{table:ion_density}.

\begin{table}
    \begin{ruledtabular}           
     \caption{Ion density as obtained with help of BRL theory. \label{table:ion_density}}   
     \begin{tabular}{l l c}          
         time$^1$ ($\mu$s)   & Fig.~\ref{fig:probe_characteristic_30Pa}, N$_i$(1/cm$^3$) & Fig.~\ref{fig:probe_characteristic_60Pa}, N$_i$(1/cm$^3$) \\
        \hline
         1 & $7.5\times 10^8$ & $7\times 10^8$ \\
         2 & $5\times 10^8$   & $3.5\times 10^8$ \\
         3 & $3\times 10^8$   & $2 \times 10^8$ \\
     \end{tabular}
     \begin{flushleft}
       $^1$Time is counted from the UV pulse start.
     \end{flushleft}

    \end{ruledtabular}
\end{table}
\begin{figure}[!h]
    \includegraphics[width=\columnwidth]{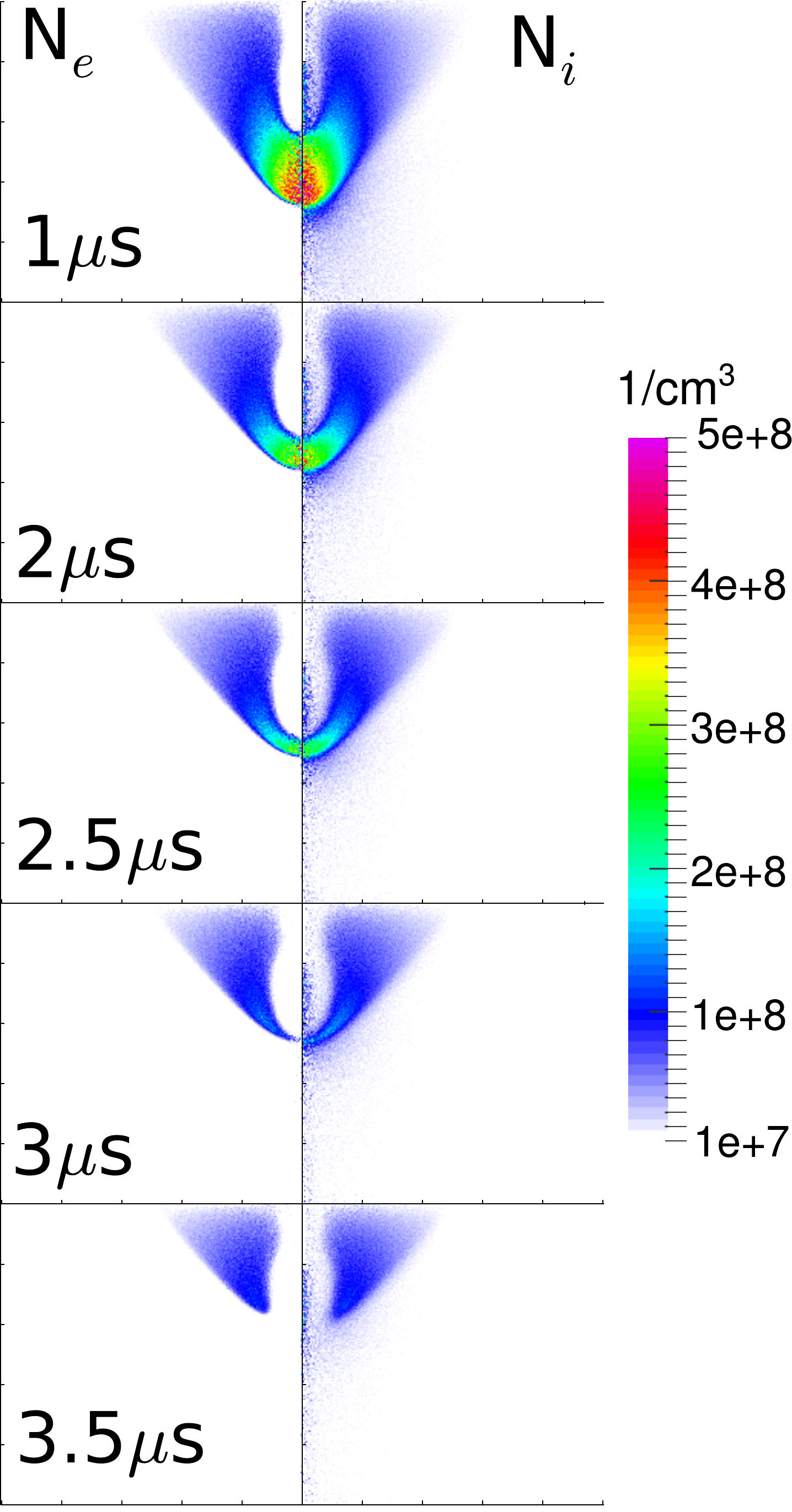}
    \caption{Electrons and ions (summed over ions types) densities for the different moments of time from UV pulse start. Simulation for 30~Pa H$_2$, -200~V bias between electrodes and -32~V on the probe. \label{fig:plasma_with_probe}}
\end{figure}

The electron temperature in the simulations after  1~$\mu$s was approximately $\sim0.5$~eV, but the EDF has a non-Maxwellian tail due to the presence of an external bias.

The probe theory predictions for the plasma density (see Table.\ref{table:ion_density}) are only comparable to the simulated plasma density near the probe tip (see Fig.~\ref{fig:plasma_with_probe}). Along the probe length, the probe theory predictions for the plasma density vary significantly from the model predictions.

However, we argue that this difference is due to an inappropriate use of probe theory. The common assumption is that the electron contribution is easily distorted, therefore, for hydrogen, one should use an electron temperature of about 0.5~eV or less. The plasma volume  can then be estimated from the total collected charge, under the assumption that, for hydrogen, ion induced emission effects are small, therefore, the plasma only decays after the UV pulse. The radial dimension of the plasma can be estimated, as the plasma occupies a cylindrical volume, hence, the plasma radius is at least about 1.5~cm, i.e. $r \geqslant \sqrt{Q_{c}/(\pi q_e\,N_i\, d)}$ , in which $Q_i \sim 4$~nC is the collected charge (see Fig.~\ref{fig:pd_scaling_charge}) and $d=5$~cm is the distance between electrodes.  Therefore, we conclude that the ratio of radial plasma size to Debye radius is, in this case, at least 75 or more, based on an estimated plasma density of $7\times10^8$~1/cm$^3$ from probe theory. Hence, the application of probe theory is validated as long as the probe characteristics have the proper shape. However, this conclusion is  wrong, since the simulations show that the plasma density varies significantly near the probe, and, the distortions of the plasma, induced by the probe are significant  (see Fig.~\ref{fig:plasma_with_probe}). Moreover,  the plasma sheath sizes are comparable with the overall plasma size. This violates the initial assumptions used in the derivation of the probe theory applied here. Therefore, an independent method for evaluating the applicability of the probe method should be used, e.g. optical emission spectroscopy.

\section{Conclusions}
In this paper, we have reported the results of a comparison between the predictions of a 2D PIC model, and experimental measurements of a low density transient plasma. 

Our 2D PIC model of the plasma evolution takes into account the full geometry of the chamber. Therefore, it allows  other experimental data, such as discharge characteristic measurements, to be used to verify assumptions about unknown and/or poorly known parameters, and to check the applicability of the model itself on the integral parameters before analysis of the complicated plasma probe characteristics.

Our results show good agreement between time resolved simulated and measured probe characteristics for two pressures with the same parameter set. This shows that the simulated plasma parameters, such as electron and ion densities, can be relied upon, because, in our simulations, the probe was included as an electrode. Therefore, the interaction of the probe with the plasma is taken into account directly.

Our measurements also show that the measured probe characteristics have the proper shape with distinct ion and electron contributions, which would normally be taken as a sign that the parameters estimated from probe measurements would be accurate.

Nevertheless, probe theories, in the case of a low density non-stationary plasma, should be used with caution. Because, our simulation results demonstrate that, even in the case where the distortion of the plasma by the probe is obviously large (see Fig.~\ref{fig:plasma_with_probe}), the probe response appears to be normal. Therefore, the application of the probe theory may yields misleading results, which in our case, is the overestimation of the plasma density by a factor of three to seven along the surface of the probe.

Hence, in the general case, the application of probe theory cannot be justified by  probe measurements alone, and an other method should be used to verify the results of the probe method.

\begin{acknowledgments}
The authors would like to thank Tatiana Rakhimova for helpful discussion.  
This work is part of the research program “Controlling photon and plasma induced processes at EUV optical surfaces (CP3E)” of the “Stichting voor Fundamenteel Onderzoek der Materie (FOM)” which is financially supported by the Nederlandse Organisatie voor Wetenschappelijk Onderzoek (NWO).
The CP3E programme is cofinanced by Carl Zeiss SMT GmbH (Oberkochen), ASML (Veldhoven), and the AgentschapNL through the Catrene EXEPT program.
\end{acknowledgments}
\bibliographystyle{unsrtnat}
\bibliography{ref.1}

\begin{thebibliography}{23}
\providecommand{\natexlab}[1]{#1}
\providecommand{\url}[1]{\texttt{#1}}
\expandafter\ifx\csname urlstyle\endcsname\relax
  \providecommand{\doi}[1]{doi: #1}\else
  \providecommand{\doi}{doi: \begingroup \urlstyle{rm}\Url}\fi

\bibitem[Hutchinson(2005)]{Hutchinson.2005.principles}
I.~H. Hutchinson.
\newblock \emph{Principles of Plasma Diagnostics}.
\newblock Cambridge University Press, July 2005.
\newblock ISBN 9780521675741.

\bibitem[Chen(2003)]{Chen.2003.langmuir}
Francis~F. Chen.
\newblock Langmuir probe diagnostics.
\newblock In \emph{Mini-Course on Plasma Diagnostics, IEEE-ICOPS Meeting, Jeju,
  Korea}, 2003.

\bibitem[Birdsall and Langdon(1985)]{Birdsall.1985.plasma}
Charles~K Birdsall and A.~Bruce Langdon.
\newblock \emph{Plasma physics via computer simulation}.
\newblock McGraw-Hill, New York, 1985.
\newblock ISBN 0070053715 9780070053717.

\bibitem[Raizer(1991)]{Raizer.1991.gas}
Yu.P. Raizer.
\newblock \emph{Gas discharge physics}.
\newblock Springer-Verlag, Berlin; New York, 1991.
\newblock ISBN 3-540-19462-2.

\bibitem[Lide(2003)]{Lide.2003.crc}
David~R Lide.
\newblock \emph{CRC handbook of chemistry and physics, 2003-2004: a
  ready-reference book of chemical and physical data.}
\newblock CRC Press, Boca Raton, Fla., 2003.
\newblock ISBN 0849304849 9780849304842.

\bibitem[Lakits et~al.(1990)Lakits, Aumayr, Heim, and
  Winter]{Lakits.1990.threshold}
G.~Lakits, F.~Aumayr, M.~Heim, and H.~Winter.
\newblock Threshold of ion-induced kinetic electron emission from a clean metal
  surface.
\newblock \emph{Physical Review A}, 42\penalty0 (9), 1990.
\newblock \doi{10.1103/PhysRevA.42.5780}.

\bibitem[Winter et~al.(1991)Winter, Aumayr, and Lakits]{Winter.1991.recent}
H.~Winter, F.~Aumayr, and G.~Lakits.
\newblock Recent advances in understanding particle-induced electron emission
  from metal surfaces.
\newblock \emph{Nuclear Instruments and Methods in Physics Research Section B:
  Beam Interactions with Materials and Atoms}, 58\penalty0 (3):\penalty0
  301--308, 1991.
\newblock \doi{10.1016/0168-583X(91)95859-C}.

\bibitem[Kishinevsky(1973)]{Kishinevsky.1973.estimation}
L.~M. Kishinevsky.
\newblock Estimation of electron potential emission yield dependence on metal
  and ion parameters.
\newblock \emph{Radiation Effects}, 19\penalty0 (1):\penalty0 23--27, January
  1973.
\newblock ISSN 0033-7579.
\newblock \doi{10.1080/00337577308232211}.

\bibitem[Agarwala and Fort(1974)]{Agarwala.1974.work}
V~Agarwala and T~Fort.
\newblock Work function changes during low pressure oxidation of aluminum at
  room temperature.
\newblock \emph{Surface Science}, 45:\penalty0 470--482, 1974.
\newblock \doi{10.1016/0039-6028(74)90183-6}.

\bibitem[Sosa(2002)]{Sosa.2002.electron}
Edward~Delarosa Sosa.
\newblock \emph{The electron emission characteristics of aluminum, molybdenum
  and carbon nanotubes studied by field emission and photoemission}.
\newblock PhD thesis, University of North Texas, 2002.

\bibitem[Helm(1988)]{Helm.1988.measurement}
Hanspeter Helm.
\newblock Measurement of the ionization potential of triatomic hydrogen.
\newblock \emph{Physical Review A}, 38\penalty0 (7):\penalty0 3425--3429,
  October 1988.
\newblock \doi{10.1103/PhysRevA.38.3425}.

\bibitem[Winter et~al.(2001)Winter, Eder, Aumayr, Lorincik, and
  Sroubek]{Winter.2001.slow-ion}
H~Winter, H~Eder, F~Aumayr, J~Lorincik, and Z~Sroubek.
\newblock Slow-ion induced electron emission from clean metal surfaces:
  ``subthreshold kinetic emission'' and ``potential excitation of plasmons''.
\newblock \emph{Nuclear Instruments and Methods in Physics Research B},
  182:\penalty0 15--22, 2001.
\newblock \doi{10.1016/S0168-583X(01)00649-8}.

\bibitem[Bogaerts and Gijbels(2002)]{Bogaerts.2002.hybrid}
Annemie Bogaerts and Renaat Gijbels.
\newblock Hybrid monte carlo{\textemdash}fluid modeling network for an
  argon/hydrogen direct current glow discharge.
\newblock \emph{Spectrochimica Acta Part B: Atomic Spectroscopy}, 57\penalty0
  (6):\penalty0 1071--1099, June 2002.
\newblock ISSN 0584-8547.
\newblock \doi{10.1016/S0584-8547(02)00047-2}.

\bibitem[Baragiola et~al.(1979)Baragiola, Alonso, and
  Florio]{Baragiola.1979.electron}
R.~A. Baragiola, E.~V. Alonso, and A.~Oliva Florio.
\newblock Electron emission from clean metal surfaces induced by low-energy
  light ions.
\newblock \emph{Physical Review B}, 19\penalty0 (1):\penalty0 121, 1979.
\newblock \doi{10.1103/PhysRevB.19.121}.

\bibitem[Langmuir(1923)]{Langmuir.1923.effect}
Irving Langmuir.
\newblock The effect of space charge and initial velocities on the potential
  distribution and thermionic current between parallel plane electrodes.
\newblock \emph{Physical Review}, 21\penalty0 (4):\penalty0 419--435, April
  1923.
\newblock \doi{10.1103/PhysRev.21.419}.

\bibitem[Phelps(1990)]{Phelps.1990.cross}
A.~V. Phelps.
\newblock Cross sections and swarm coefficients for h+, h2+, h3+, h, h2, and h-
  in h2 for energies from 0.1 ev to 10 kev.
\newblock \emph{Journal of Physical and Chemical Reference Data}, 19\penalty0
  (3):\penalty0 653, 1990.
\newblock ISSN 00472689.
\newblock \doi{10.1063/1.555858}.

\bibitem[\v{S}imko et~al.(1997)\v{S}imko, Marti\v{s}ovit\v{s}, Bretagne, and
  Gousset]{Simko.1997.transport}
T.~\v{S}imko, V.~Marti\v{s}ovit\v{s}, J.~Bretagne, and G.~Gousset.
\newblock Computer simulations of h$^+$ and h$_3^+$ transport parameters in
  hydrogen drift tubes.
\newblock \emph{Physical Review E}, 56\penalty0 (5):\penalty0 5908--5919,
  November 1997.
\newblock \doi{10.1103/PhysRevE.56.5908}.

\bibitem[Mokrov and Raizer(2008)]{Mokrov.2008.monte_carlo}
M.~S. Mokrov and Yu~P. Raizer.
\newblock Monte carlo method for finding the ionization and secondary emission
  coefficients and i{\textendash}v characteristic of a townsend discharge in
  hydrogen.
\newblock \emph{Technical Physics}, 53\penalty0 (4):\penalty0 436--444, April
  2008.
\newblock ISSN 1063-7842, 1090-6525.
\newblock \doi{10.1134/S1063784208040075}.

\bibitem[Nanbu(1994)]{Nanbu.1994.simple}
Kenichi Nanbu.
\newblock Simple method to determine collisional event in monte carlo
  simulation of electron-molecule collision.
\newblock \emph{Japanese Journal of Applied Physics}, 33\penalty0 (Part 1, No.
  8):\penalty0 4752--4753, August 1994.
\newblock ISSN 0021-4922.
\newblock \doi{10.1143/JJAP.33.4752}.

\bibitem[Skullerud(1968)]{Skullerud.1968.stochastic}
H.~R. Skullerud.
\newblock The stochastic computer simulation of ion motion in a gas subjected
  to a constant electric field.
\newblock \emph{Journal of Physics D: Applied Physics}, 1\penalty0
  (11):\penalty0 1567, 1968.
\newblock \doi{10.1088/0022-3727/1/11/423}.

\bibitem[Dutton(1975)]{Dutton.1975.survey}
J.~Dutton.
\newblock A survey of electron swarm data.
\newblock \emph{Journal of Physical and Chemical Reference Data}, 4\penalty0
  (3):\penalty0 577--856, July 1975.
\newblock ISSN 0047-2689, 1529-7845.
\newblock \doi{10.1063/1.555525}.

\bibitem[Graham et~al.(1973)Graham, James, Keever, Albritton, and
  McDaniel]{Graham.1973.mobilities}
E.~Graham, D.~R. James, W.~C. Keever, D.~L. Albritton, and E.~W. McDaniel.
\newblock Mobilities and longitudinal diffusion coefficients of mass-identified
  hydrogen ions in h2 and deuterium ions in d2 gas.
\newblock \emph{The Journal of Chemical Physics}, 59\penalty0 (7):\penalty0
  3477--3481, October 1973.
\newblock ISSN 00219606.
\newblock \doi{doi:10.1063/1.1680505}.

\bibitem[Laframboise(1966)]{Laframboise.1966.theory}
James~G. Laframboise.
\newblock Theory of spherical and cylindrical langmuir probes in a
  collisionless, maxwellian plasma at rest.
\newblock Technical Report UTIAS Report NO. 100, DTIC Document, University of
  Toronto, Institute for Aerospace studies, 1966.

\end{thebibliography}
\end{document}